\documentclass[reprint, superscriptaddress, secnumarabic, amssymb, nobibnotes, aps, prb]{revtex4-2}

\usepackage{graphicx}
\usepackage{epstopdf}
\usepackage[T1]{fontenc}
\usepackage[latin9]{inputenc}
\usepackage{amsbsy}
\usepackage{gensymb}
\usepackage[T1]{fontenc}
\usepackage[latin9]{inputenc}
\usepackage{amsmath}
\usepackage{amssymb}
\usepackage{bbm}
\usepackage{physics}
\usepackage{braket}
\usepackage{xcolor}
\allowdisplaybreaks
\usepackage{graphicx}
\usepackage[colorlinks=true]{hyperref}  
\hypersetup{
    bookmarks=true,         
    unicode=false,          
    pdftoolbar=true,        
    pdfmenubar=true,        
    pdffitwindow=false,     
    pdfstartview={FitH},    
    pdftitle={cenige3},    
    pdfauthor={},     
    pdfsubject={},   
    pdfcreator={},   
    pdfproducer={}, 
    pdfkeywords={} {} {}, 
    pdfnewwindow=true,      
    colorlinks=true,       
    linkcolor=blue, 
    citecolor=blue,        
    filecolor=magenta,      
    urlcolor=blue           
} 
\usepackage[normalem]{ulem}

\newcommand{\equref}[1]{Eq.~(\ref{#1})}

\newcommand{\figref}[1]{Fig.~\ref{#1}}
\newcommand{\refcite}[1]{Ref.~\onlinecite{#1}}

\renewcommand{\approx}{\simeq}
\renewcommand{\Re}{\text{Re}}

\begin{document}

\title{\textrm{Magnetic structure and crystal field states of antiferromagnetic CeNiGe$_3$: Neutron scattering and
$\mu$SR investigations}}
\author{A. Kataria}
\altaffiliation{anshuk@jncasr.ac.in}
\affiliation{ISIS Neutron and Muon Source, Rutherford Appleton Laboratory, Chilton, Didcot, Oxon, OX11 0QX, United Kingdom}
\affiliation{School of Advanced Materials and Chemistry and Physics of Materials Unit, Jawaharlal Nehru Centre for Advanced Scientific Research, Jakkur, Bangalore 560064, India}
\author{R. Kumar}
\affiliation{School of Advanced Materials and Chemistry and Physics of Materials Unit, Jawaharlal Nehru Centre for Advanced Scientific Research, Jakkur, Bangalore 560064, India}
\author{D.~T.~Adroja}
\altaffiliation{devashibhai.adroja@stfc.ac.uk}
\affiliation{ISIS Neutron and Muon Source, Rutherford Appleton Laboratory, Chilton, Didcot, Oxon, OX11 0QX, United Kingdom}
\affiliation {Highly Correlated Matter Research Group, Physics Department, University of Johannesburg, P.O. Box 524, Auckland Park 2006, South Africa}
\author{C. Ritter}
\affiliation{Institut Laue-Langevin, Boite Postale 156, 38042 Grenoble Cedex, France}
\author{V.~K.~Anand}
\affiliation{ISIS Neutron and Muon Source, Rutherford Appleton Laboratory, Chilton, Didcot, Oxon, OX11 0QX, United Kingdom}
\affiliation{Department of Mathematics and Physics, University of Stavanger, 4036 Stavanger, Norway}
\affiliation{Department of Physics, National Institute of Technology Agartala, Tripura 799046, India}
\author{A. D. Hillier}
\affiliation{ISIS Neutron and Muon Source, Rutherford Appleton Laboratory, Chilton, Didcot, Oxon, OX11 0QX, United Kingdom}
\author{B. M. Huddart} 
\affiliation{Department of Physics, Centre for Materials Physics, Durham University, Durham, DH1 3LE, United Kingdom}
\affiliation{Clarendon Laboratory, University of Oxford, Department of Physics, Oxford OX1 3PU, United Kingdom}
\author{T. Lancaster}
\affiliation{Department of Physics, Centre for Materials Physics, Durham University, Durham, DH1 3LE, United Kingdom}

\author{S. Rols}
\affiliation{Institut Laue-Langevin, Boite Postale 156, 38042 Grenoble Cedex, France}
\author{M. M. Koza}
\affiliation{Institut Laue-Langevin, Boite Postale 156, 38042 Grenoble Cedex, France}
\author{Sean Langridge}
\affiliation{ISIS Neutron and Muon Source, Rutherford Appleton Laboratory, Chilton, Didcot, Oxon, OX11 0QX, United Kingdom}
\author{ A. Sundaresan}
\affiliation{School of Advanced Materials and Chemistry and Physics of Materials Unit, Jawaharlal Nehru Centre for Advanced Scientific Research, Jakkur, Bangalore 560064, India}
\begin{abstract}

We present the results of microscopic investigations of antiferromagnetic CeNiGe$_3$, using neutron powder diffraction (NPD), inelastic neutron scattering (INS), and muon spin relaxation ($\mu$SR) measurements. CeNiGe$_3$ crystallizes in a centrosymmetric orthorhombic crystal structure (space group: $Cmmm$) and undergoes antiferromagnetic (AFM) ordering. The occurrence of long-range AFM ordering at $T_{\rm N} \approx 5.2$~K is confirmed by magnetic susceptibility, heat capacity, neutron diffraction, and $\mu$SR measurements. The NPD data characterize the AFM state with an incommensurate helical magnetic structure having a propagation vector \textbf{k} = (0, 0.41, 1/2). In addition, INS measurements at 10~K identified two crystal electric field (CEF) excitations at 9.17~meV and 18.42~meV. We analyzed the INS data using a CEF model for an orthorhombic environment of Ce$^{3+}$ ($J=5/2$) and determined the CEF parameters and ground state wavefunctions of CeNiGe$_3$. Moreover, zero-field $\mu$SR data for CeNiGe$_3$ at $T< T_{\rm N}$ show long-range AFM ordering with three distinct oscillation frequencies corresponding to three different internal fields at the muon sites. The internal fields at the muon-stopping sites have been further investigated using density functional theory  calculations. 

\end{abstract}

\date{\today}

\maketitle

\section{Introduction}

Rare-earth intermetallics have gained significant research attention due to their exotic physical properties arising from strong electronic correlations. In particular, Ce-based systems exhibit competing inter-site exchange and on-site Kondo interactions between localised $f$-moments and conduction electrons, prompting unusual ground states and properties, including valence fluctuations, unconventional superconductivity, heavy fermion behavior and quantum criticality \cite{hf, hf2, hf3,vf, hf_sc,hf_sc1, qc, qc1, qc2}. The electronic ground state and physical properties of these systems primarily depend on the competition between long-range Ruderman-Kittel-Kasuya-Yosida (RKKY) and on-site Kondo interactions as well as on the ground state crystal electric field (CEF) wave functions. 

Ce-based intermetallic systems Ce$TX_3$ ($T$ = transition metal and $X$ = Si, Ge) with a BaNiSn$_3$-type tetragonal structure (space group $I4mm$) have been widely investigated owing to the lack of inversion symmetry and strong CEF-phonon coupling  \cite{Cecoge3_muon, Ceirge3_muon, cef_coupling,Cecuge3_muon,mmt,mmt2, Ceramak2019, Hillier2012, CTX3_SC, CTX3_SC2, Cerhsi3, Cerhsi3_sc, CeIrge3_sc,ceirsi3_sc, Cecoge3_sc}. In the absence of inversion symmetry, the pressure-induced superconductivity in Ce$TX_3$ is suggested to exhibit an admixture state of spin-singlet and spin-triplet parity, caused by antisymmetric spin-orbit coupling \cite{CTX3_SC, CTX3_SC2, Cerhsi3, Cerhsi3_sc, CeIrge3_sc,ceirsi3_sc, Cecoge3_sc}. Further, according to Kramers' degeneracy theorem, Ce$^{3+}$ being a Kramers ion should have only two CEF excitations from the ground state doublet; however, the experiments reveal three CEF excitations in CeCuAl$_3$ \cite{cef_coupling}, CeAuAl$_3$ \cite{Ceramak2019} and CeCuGa$_3$ \cite{Cecuge3_muon}, which showcase the strong CEF-phonon coupling and interesting interplay between phononic and CEF interactions.  

While the majority of Ce$TX_3$ compounds crystallize in the noncentrosymmetric BaNiSn$_3$-type tetragonal structure (space group $I4mm$), they also adopt different crystal structures such as a SmNiGe$_3$-type orthorhombic structure ($Cmmm$), CePtGa$_3$-type orthorhombic ($ Fmm2$), CeNiSb$_3$-type orthorhombic ($Pbcm$), LaRuSn$_3$-type cubic quasi-skutterudite structure ($Pm$-$3n$), BaNiO$_3$-type hexagonal structure ($P6_3/mmc$) as well as their superstructure variants \cite{rev_stru, CeRuGe3, Anand2016, Anand2011, Das2015, CNG3_stru,CNG3_poly}. The present work focuses on antiferromagnetic CeNiGe$_3$, which crystallizes in a SmNiGe$_3$-type centrosymmetric orthorhombic structure \cite{CNG3_stru,CNG3_poly}. Similar to the BaNiSn$_3$-type, the SmNiGe$_3$-type structure exhibits layers of Ge$_2$Ni-SmGe$_2$Sm-NiGe$_2$ stacked along the $b$-axis. In CeNiGe$_3$, Ce and one Ge atom occupy Wyckoff sites  $4j$ ($x$, 0, 1/2), whereas Ni and the remaining two Ge atoms are placed on Wyckoff sites $4i$ ($x$, 0, 0) with different $x$; see Table~\ref{tab:NPD} for details.

Earlier investigations on polycrystalline CeNiGe$_3$ by Pikul {\it et al}. \cite{CNG3_poly} report antiferromagnetic (AFM) ordering below 5.5~K, whereas Durivault {\it et al}. \cite{Durivault2003} observed two magnetic transitions at 5.9 K and 5.0 K. The neutron powder diffraction (NPD) study by Durivault {\it et al}. \cite{Durivault2003} revealed a collinear AFM structure [wavevector {\bf k$_1$} = (1,0,0)] below 5.9~K, which coexists with an incommensurate helicoidal magnetic structure [{\bf k$_2$} = (0,0.409(1),1/2)] below 5.0~K, that is largely preponderant at the lowest temperature \cite{Durivault2003}. In contrast, susceptibility data on single crystalline CeNiGe$_3$ report the onset of a single AFM ordering near 5.0~K \cite{CNG3_SC}, and NPD for a single crystal also found a single incommensurate wave vector {\bf k$_2$} = (0,0.41,1/2) \cite{CNG3_singlecrystal}.  The nuclear quadrupolar resonance (NQR) results on a polycrystalline CeNiGe$_3$ also supports the presence of a single incommensurate magnetic structure at 5.1~K \cite{nqr1,nqr2}.
 
CeNiGe$_3$ exhibits intriguing field-temperature $(H$-$T)$ and pressure-temperature $(P$-$T)$ phase diagrams \cite{CNG3_SC, CNG3_poly}. A pressure study on CeNiGe$_3$ reveals two superconducting (SC) domes \cite{CNG3_scp, CNG3_scp2, CNG3_scp3, nqr1}. The superconductivity appears at pressures above 2.3 GPa. This SC phase (SC-I) coexists with a first antiferromagnetic (AFM-I) phase where a further increase in pressure suppresses the SC phase at 3.7 GPa and causes a phase transition to a second AFM-II phase \cite{CNG3_scp3,nqr1,nqr2}. SC reappears (SC-II phase) over a narrow pressure range as the pressure further increases. Interestingly, in the AFM-I phase, the ordering temperature increases with increasing pressure, while in the AFM-II phase, the ordering temperature decreases with increasing pressure and is completely suppressed at 6.5 GPa \cite{CNG3_scp3}. The pressure dependence of the AFM-I phase differs from the quantum critical behavior usually observed in Ce-based compounds. Further, the field-tuning behavior of the AFM state also does not reflect quantum critical behavior \cite{CNG3_SC}.   

A recent femtosecond-resolved coherent phonon spectroscopy study reports the absence of the Kondo effect in CeNiGe$_3$ \cite{CNG3_phn}. This finding is counter-intuitive to the previously believed Kondo lattice heavy fermion-like scenario based on electrical resistivity and heat capacity data for CeNiGe$_3$ \cite{CNG3_poly}, and reflects a weak Kondo effect in this compound. The coherent phonon dynamics are explained by an anharmonic phonon-phonon scattering effect and suggest a mode splitting around 105 K, which is close to the CEF splitting energy of 100 K between the ground state and first excited state, as inferred from the CEF analysis of magnetic susceptibility $\chi$ and heat capacity $C_{\rm p}$ data of single-crystalline CeNiGe$_3$ \cite{CNG3_phn, CNG3_singlecrystal}. In the same study, it is proposed that thermal energy can change the energy of hybridization, and at high temperatures, the conduction band hybridizes with the first excited CEF level, leading to an orbital crossover causing mode splitting in CeNiGe$_3$ \cite{CNG3_phn}. The CEF thus has a significant effect on lattice vibrations and electron occupancy, and since the Kondo effect is highly dependent on the orientations and electronic occupancies of different CEF orbitals, the CEF plays a critical role in the absence of the Kondo effect in CeNiGe$_3$. This motivated us to investigate the CEF effect in CeNiGe$_3$ using inelastic neutron scattering (INS) measurements, which can help us understand the role of CEF and its impact on the ground state properties. 

We have investigated the magnetism and crystal field in CeNiGe$_3$ using the NPD, INS, and muon spin relaxation ($\mu$SR)  measurements on a polycrystalline sample. The $\chi(T)$ and $C_{\rm p}(T)$ measurements were performed for comparison with the literature results. Our sample shows a single antiferromagnetic transition near 5.2~K, and NPD data reveal an incommensurate AFM structure with \textbf{k} = (0, 0.4088(1), 1/2), consistent with the previous report on a single crystalline sample \cite{CNG3_singlecrystal}. In addition, the INS measurements at 10 K and 100 K reveal the presence of two CEF excitations from $J$ = 5/2 ground state multiplet of Ce$^{3+}$ ions, which are characteristic of the transition between the ground state and the excited Kramers doublets states. The $\mu$SR study provides additional confirmation of bulk magnetic ordering below 5.27(6) K with multiple internal fields at the muon-stopping site. Density functional theory (DFT)  calculations have been performed to identify the muon-stopping sites and estimate the internal fields.

\section{Experimental Details}

The polycrystalline samples of CeNiGe$_3$ and its nonmagnetic analogue YNiGe$_3$ were prepared by arc-melting the constituting high purity (99.9\% and above) elements in stoichiometric ratio under an argon atmosphere and subsequent annealing at $900^\circ$C for a week. The quality of the CeNiGe$_3$ and YNiGe$_3$ samples prepared were checked by room temperature powder x-ray diffraction, revealing the samples to be good quality and in single-phase. The temperature $T$ and field $H$ dependent magnetization $M$ measurements were performed using a SQUID, Quantum Design Magnetic Property Measurement System (QD-MPMS), while the heat capacity was measured using the Quantum Design Physical Property Measuring System (QD-PPMS). The NPD data were collected on the high-intensity diffractometer D1B at the Institut Laue Langevin (ILL) in Grenoble, France. About 20 g of CeNiGe$_3$ powder sample was placed in a cylindrical Vanadium can and cooled using a standard orange cryostat. Long measurements of 4.5 hours each were recorded at 1.5 K and 5.5 K. A thermodiffractogram was measured between 2.5 K and 6.7 K increasing the temperature with a ramp speed of 0.01 K/60 secs. Data were taken for 10 minutes resulting in a temperature resolution of 0.1 K between adjacent data points. INS measurements were carried out using the IN4C time-of-flight spectrometer as well at ILL, with incident energy of $E_i$ = 18 and 41.7~meV. INS measurements were also conducted on the nonmagnetic reference YNiGe$_3$ to estimate the phonon contribution and extract the magnetic contribution of the CeNiGe$_3$ sample. The samples were mounted in an Al-foil, placed in a flat plate-type Al-holder and cooled in an orange cryostat to the lowest temperature of 10 K.

\begin{figure*}
\centering
    \includegraphics[width=\linewidth]{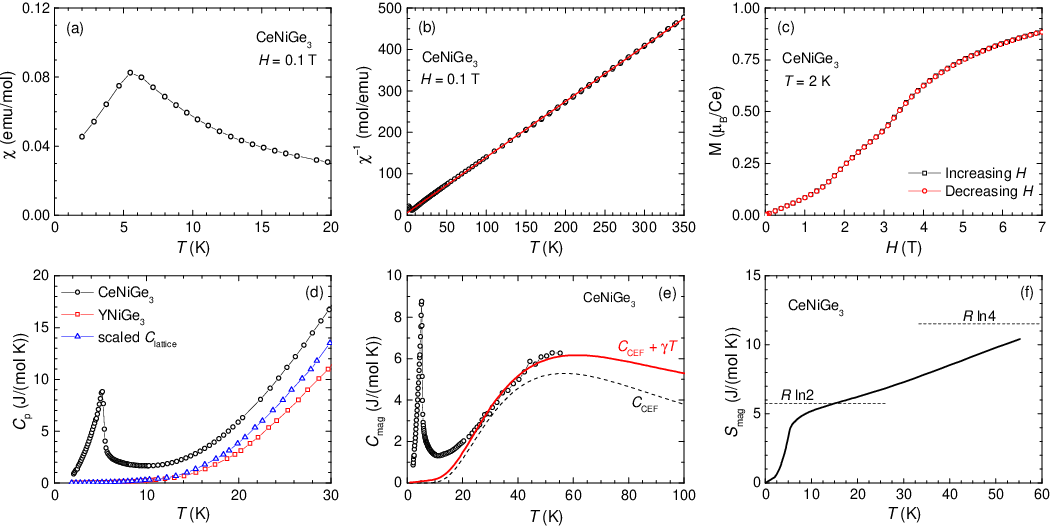}
    \caption{\label{Fig1:charc} (a) Temperature $T$ variation of magnetic susceptibility $\chi$ of CeNiGe$_3$ for $2 \leq T \leq20$~K measured in a magnetic field 
       $H =0.1$~T\@. (b) Inverse magnetic susceptibility $\chi^{-1}(T)$ for $2 \leq T \leq 350$~K measured in $H =0.1$~T. The solid red line represents the Curie-Weiss law fitting. (c) Isothermal $M$ vs $H$ loop at 2~K. (d) Heat capacity $C_{\rm p}(T)$ of CeNiGe$_3$ measured in zero field along with the $C_{\rm p}(T)$ of non-magnetic counterpart YNiGe$_3$ from Ref.~\cite{CNG3_poly} which was used to obtain the mass corrected lattice contribution to heat capacity $C_{\rm lattice}(T)$. (e) Magnetic contribution to heat capacity $C_{\rm mag}(T)$. The solid red curve represents the crystal electric field heat capacity $C_{\rm CEF}(T)$ (obtained from the analysis of the inelastic neutron scattering data) plus an electronic contribution $\gamma T$ for $\gamma = 15$~mJ/mol\, K$^2$. (f) Entropy from Magnetic contribution $S_{\rm mag}(T)$. }
\end{figure*}

Zero-field (ZF)-$\mu$SR measurements were performed on the MuSR spectrometer at ISIS, UK. Three sets of orthogonal coils have been arranged in such a way to ensure that there is no stray magnetic field at the sample's position. More details about the MuSR instrument and methodology can be found in \refcite{muon, muon2, muonbook}. The base temperature of 1.2~K was achieved using a He-4 cryostat with pumping on the He-bath.   The crystal and magnetic structure refinement of NPD data was performed using FullProf \cite{fullprof} and Jana software \cite{jana}. The INS and $\mu$SR asymmetry spectra have been analyzed using MANTID and WiMDA software, respectively \cite{mantid,wimda}.

\section{Results and Discussion}
\subsection {Magnetic Susceptibility and Heat Capacity}

The low-temperature magnetic susceptibility of the polycrystalline CeNiGe$_3$ measured in a magnetic field $H =0.1$~T over $2 \leq T \leq20$~K is presented in \figref{Fig1:charc}(a). A peak at low $T$ in $\chi(T)$ reveals  antiferromagnetic ordering at a transition temperature $T_{\rm N}$ = 5.5 K (see \figref{Fig1:charc}(a)), which is consistent with a previous report \cite {CNG3_poly}. The inverse magnetic susceptibility $\chi^{-1}(T)$ measured in $H =0.1$~T is shown in \figref{Fig1:charc}(b) for $2 \leq T \leq 350$~K\@. The Curie-Weiss law, $\chi= C/(T-\theta_p)$, is used to fit $\chi^{-1}(T)$ in the range 50 K to 350 K. The best fit to the data yields a Weiss temperature, $\theta_{\rm p} = -7.7(6)$~K and an effective moment, $\mu_{\mathrm{eff}}$ = 2.44(1) $\mu_\mathrm{B}$, which is close to the theoretical value for a free Ce$^{3+}$ ion, $\mu_{\mathrm{eff}}= $ 2.54 $\mu_\mathrm{B}$. The observed $\mu_{\mathrm{eff}}$ closely aligns with the reported values for both the polycrystalline average of a single-crystalline sample (2.57 $\mu_\mathrm{B}$) and the polycrystal sample (2.47 $\mu_\mathrm{B}$) \cite{CNG3_singlecrystal, CNG3_poly}. However, our $\theta_{\rm p} = -7.7(6)$~K is significantly lower than the reported values of $-21$~K for single crystal \cite{CNG3_singlecrystal} and $-17$~K for polycrystalline sample \cite{CNG3_poly}, where the negative sign indicates a dominant antiferromagnetic interaction for this system. 

Figure~\ref{Fig1:charc}(c) shows the isothermal magnetization $M(H)$ measured at 2 K, indicating a non-linear, high-field saturation behaviour with no hysteresis. The $M$ vs $H$ curve displays two distinct humps with increasing fields associated with meta-magnetic transitions at 1.75(1) T and 3.27(2) T, respectively. The observed behaviour and obtained values are in close agreement with reported results for both single-crystal (1.85 T and 3.10 T at 2.0 K) and polycrystalline samples (1.72 T and 3.25 T at 1.7~K) \cite{CNG3_singlecrystal, CNG3_poly}. Even at 7 T, the 2 K $M(H)$ does not saturate and attains a value of only 0.9 $\mu_\mathrm{B}$, which is little higher than the reported value 0.6 $\mu_\mathrm{B}$ for the polycrystalline sample at 5 T \cite{CNG3_poly}. A value of 1.65 $\mu_\mathrm{B}$ was reported for a single-crystalline sample \cite{CNG3_singlecrystal} for a field of 7 T applied along the crystal $a$-axis while for the field along $b$ or $c$ the saturation amounted only to about 0.15 $\mu_\mathrm{B}$ reflecting a strong anisotropy. These values are significantly lower than the theoretically expected saturation value based on Hund's rule for Ce$^{3+}$ ions ($J =$ 5/2), i.e., 2.14 $\mu_\mathrm{B}$. In the absence of any Kondo effect, this reduced moment can be attributed to the crystal electric field.

The temperature-dependent heat capacity data of CeNiGe$_3$ measured in zero-field is shown in \figref{Fig1:charc}(d) for $2 \leq T \leq 30$~K\@. A $\lambda$-like anomaly in $C_{\rm p}(T)$ with a peak at 5.2~K confirms the occurrence of antiferromagnetic ordering. Our $C_{\rm p}(T)$ data are in very good agreement with the literature data \cite{CNG3_singlecrystal, CNG3_poly}. A linear fit of the $C_{\rm p}/T$ vs $T^2$ plot from 15~K to 25~K according to $C_{\rm p}/T = \gamma + \beta T^2$ yields an electronic coefficient $ \gamma = 33(2)$~mJ/mol\,K$^2$ and a phononic coefficient $\beta = 0.65(1)$~mJ/mol\,K$^4$. The value of $ \gamma$ obtained does not reflect a heavy fermion behavior in CeNiGe$_3$ and is consistent with the weak or negligible Kondo effect. The value of $\beta$ corresponds to a Debye temperature $\Theta_{\rm D} = 246(5)$~K\@.

In order to estimate the magnetic contribution to the heat capacity $C_{\rm mag}(T)$ of CeNiGe$_3$, we use the $C_{\rm p}(T)$ of its nonmagnetic counterpart YNiGe$_3$ (which is isostructural to CeNiGe$_3$) for the estimate of the phononic contribution. The $C_{\rm p}(T)$ data of YNiGe$_3$ taken from Ref.~\cite{CNG3_poly} is shown in \figref{Fig1:charc}(d). However, since CeNiGe$_3$ and YNiGe$_3$  have different masses, a correction for the mass difference was made following Ref.~\cite{Anand2015} to obtain the phonon contribution to the heat capacity. The mass-corrected lattice contribution $C_{\rm lattice}(T)$ is shown in \figref{Fig1:charc}(d). The $C_{\rm mag}(T)$ data, obtained after subtracting  $C_{\rm lattice}(T)$ from $C_{\rm p}(T)$ for CeNiGe$_3$, are presented in \figref{Fig1:charc}(e) and compared with the crystal electric field contribution to the heat capacity $C_{\rm CEF}(T)$ (dashed line).  The $C_{\rm CEF}(T)$ data were estimated using the CEF parameters obtained from the analysis of the inelastic neutron scattering data (discussed later). The $C_{\rm mag}(T)$ data present the broad Schottky-type anomaly feature as reported previously for both polycrystalline and single crystalline samples of CeNiGe$_3$ \cite{CNG3_singlecrystal, CNG3_poly}. 
It is seen that $C_{\rm mag}(T)$ is higher than $C_{\rm CEF}(T)$ and an electronic contribution $\gamma T$ (with $\gamma = 15$~mJ/mol\,K$^2$) is required to obtain a reasonable agreement between the experimental $C_{\rm mag}(T)$ and calculated $C_{\rm CEF}(T)$ [see \figref{Fig1:charc}(e) for the comparison of $C_{\rm mag}(T)$ and $C_{\rm CEF}(T) + \gamma T$ (solid red curve)]. The need for a $\gamma T$ contribution can be attributed to the difference in $\gamma$ values of CeNiGe$_3$ and YNiGe$_3$.  The difference in the density of states at the Fermi level and the complexities in the Fermi surfaces of CeNiGe\textsubscript{3} as well as a weak contribution from the spin degree of freedom from Ce-4\textit{f}\textsuperscript{1} electrons could be the possible cause behind the additional $\gamma$. Pikul {\it et al}. \cite{CNG3_poly} suggested that a $\gamma T$ contribution with $\gamma = 45$~mJ/mol\,K$^2$ was required in their CEF analysis of $C_{\rm mag}(T)$ data. Apparently, they needed an even higher electronic contribution, as they did not account for the mass differences of CeNiGe$_3$ and YNiGe$_3$. 

The temperature-dependence of the magnetic contribution to the entropy $S_{\rm mag}(T)$ obtained by integrating the $C_{\rm mag}/T$ vs $T$ plot is shown in \figref{Fig1:charc}(f). It is seen that at $T_{\rm N}$, the $S_{\rm mag}$ attains a much lower value of $\sim 0.7 \,R \ln 2 $ than the expected $ R \ln 2 $ value for a doublet ground state. A value of $0.9 \,R \ln 2 $ is attained at 10~K, and the value of $R \ln 2 $ is achieved near 15~K\@. The reduced value of $S_{\rm mag}$ can be attributed to the development of short-range magnetic correlations well above the occurrence of long-range magnetic ordering at $T_{\rm N}$. 

\begin{figure}
\centering
    \includegraphics[width=\linewidth]{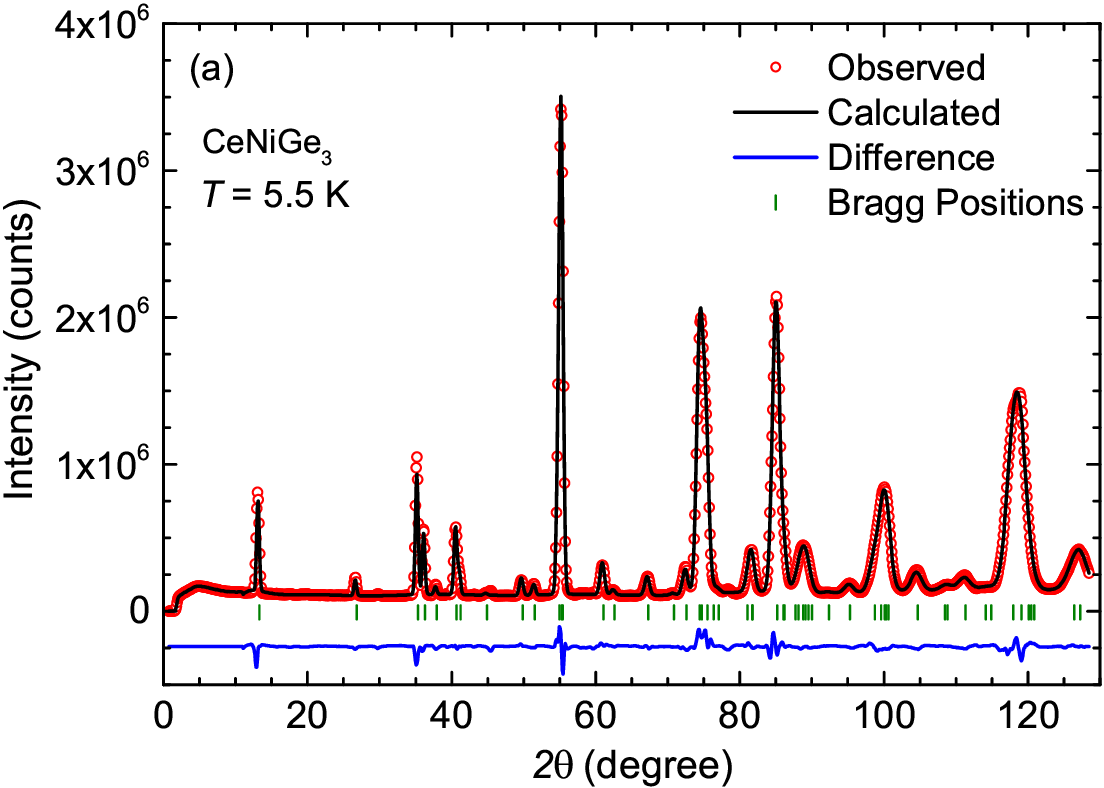} \vspace{0.1cm}
    \includegraphics[width=\linewidth]{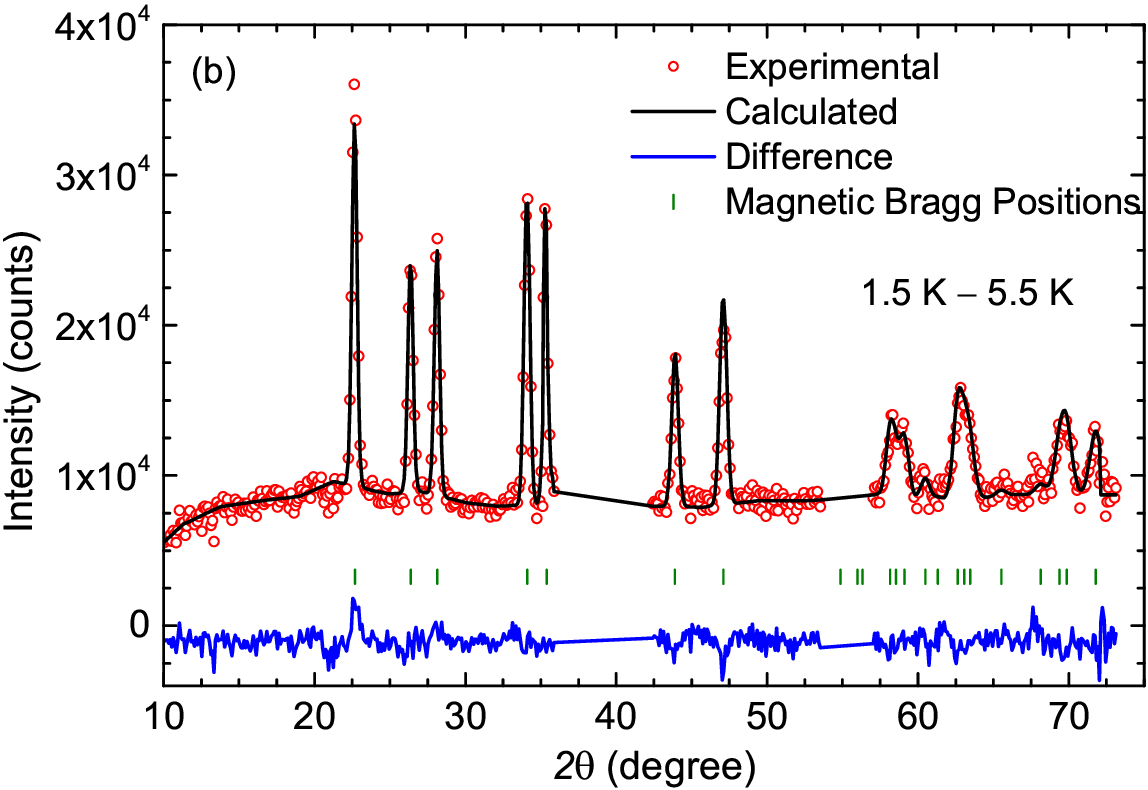}
    \caption{\label{Fig3:neutron diff} (a) Neutron powder diffraction (NPD) pattern of CeNiGe$_3$ at 5.5 K along with the nuclear refinement measured on the D1B diffractometer using wavelength of 2.51 \AA. (b) Magnetic only pattern of CeNiGe$_3$ at 1.5 K obtained after subtracting the 5.5~K NPD data from 1.5~K NPD data along with the magnetic refinement. The intermediate portions, which have contributions from the nuclear Bragg peaks and sample holder, have been removed.} 
\end{figure}

\begin{figure}
\centering
     \includegraphics[width=\columnwidth]{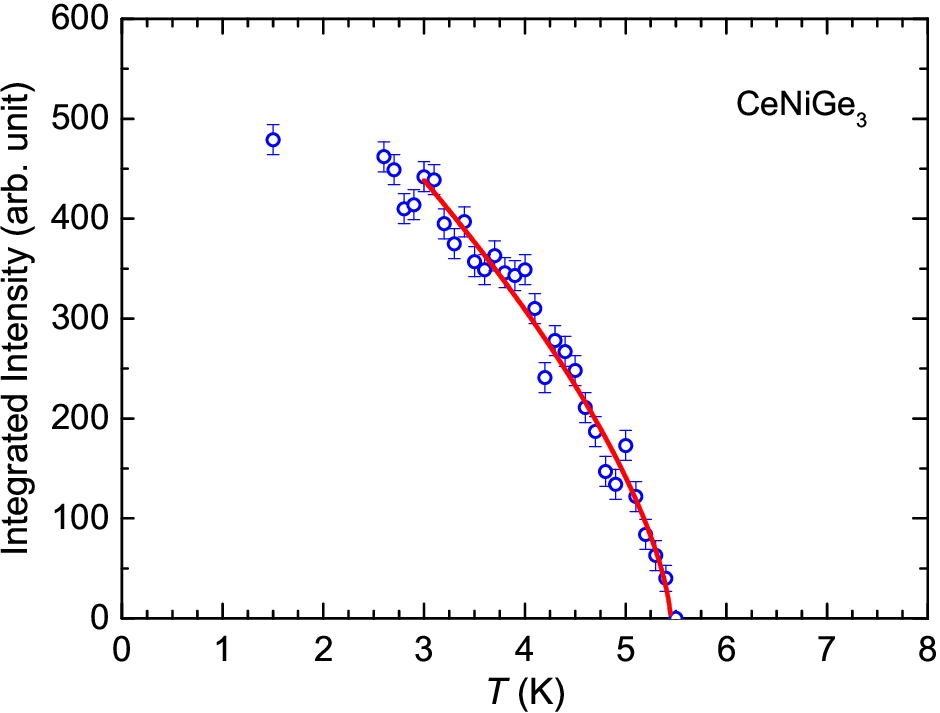}
    \caption{\label{Fig:Mag_int} Determination of the magnetic transition temperature of CeNiGe$_3$ from the temperature dependence of the integrated intensity of the strongest magnetic peak at 2$\theta$ = 22.68~(degree), with ($h$ $ k$ $l$) = (0, 0.41, 1/2),  as obtained using the thermodiffractogram. Solid red curve represents the fit to the critical behavior $I = I_0(1 - T /T_{\rm N})^{2 \beta}$ for $3.0~{\rm K} \leq T \leq T_{\rm N}$.}
\end{figure}

\begin{figure}
\centering
     \includegraphics[width=\columnwidth]{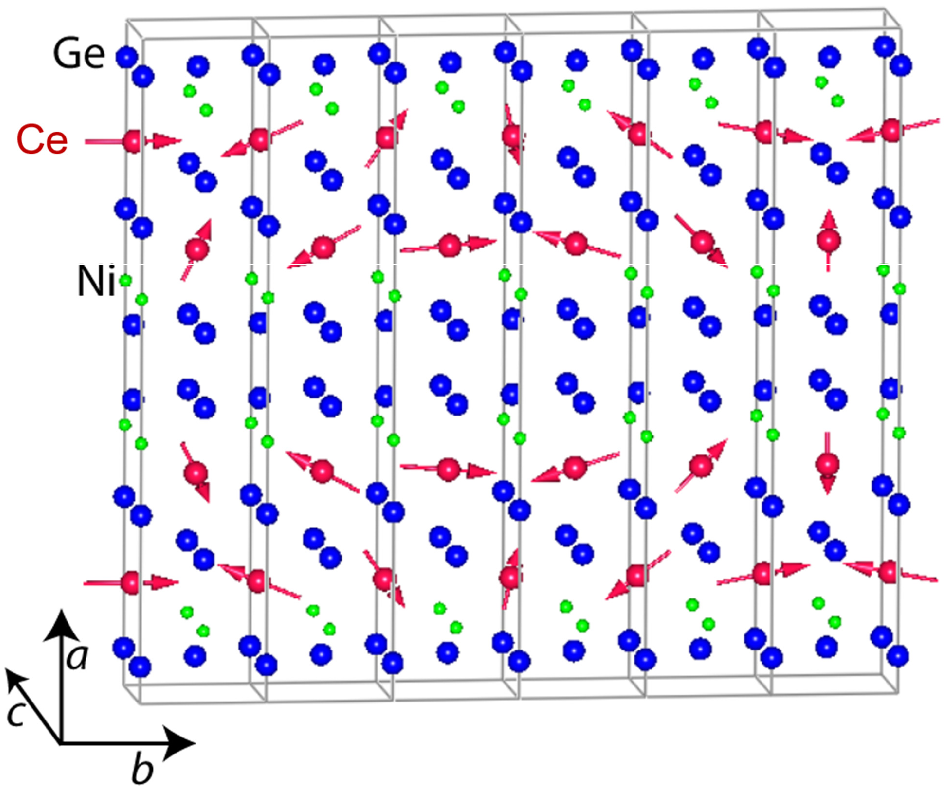}
    \caption{\label{Fig:Mag_Struct} Incommensurate helicoidal magnetic structure of CeNiGe$_3$ obtained from the refinement of neutron diffraction data. The magnetic spins are confined and rotate within the $a$$b$ plane.}
\end{figure}

\begin{figure*}
\centering
    \includegraphics[width=\linewidth]{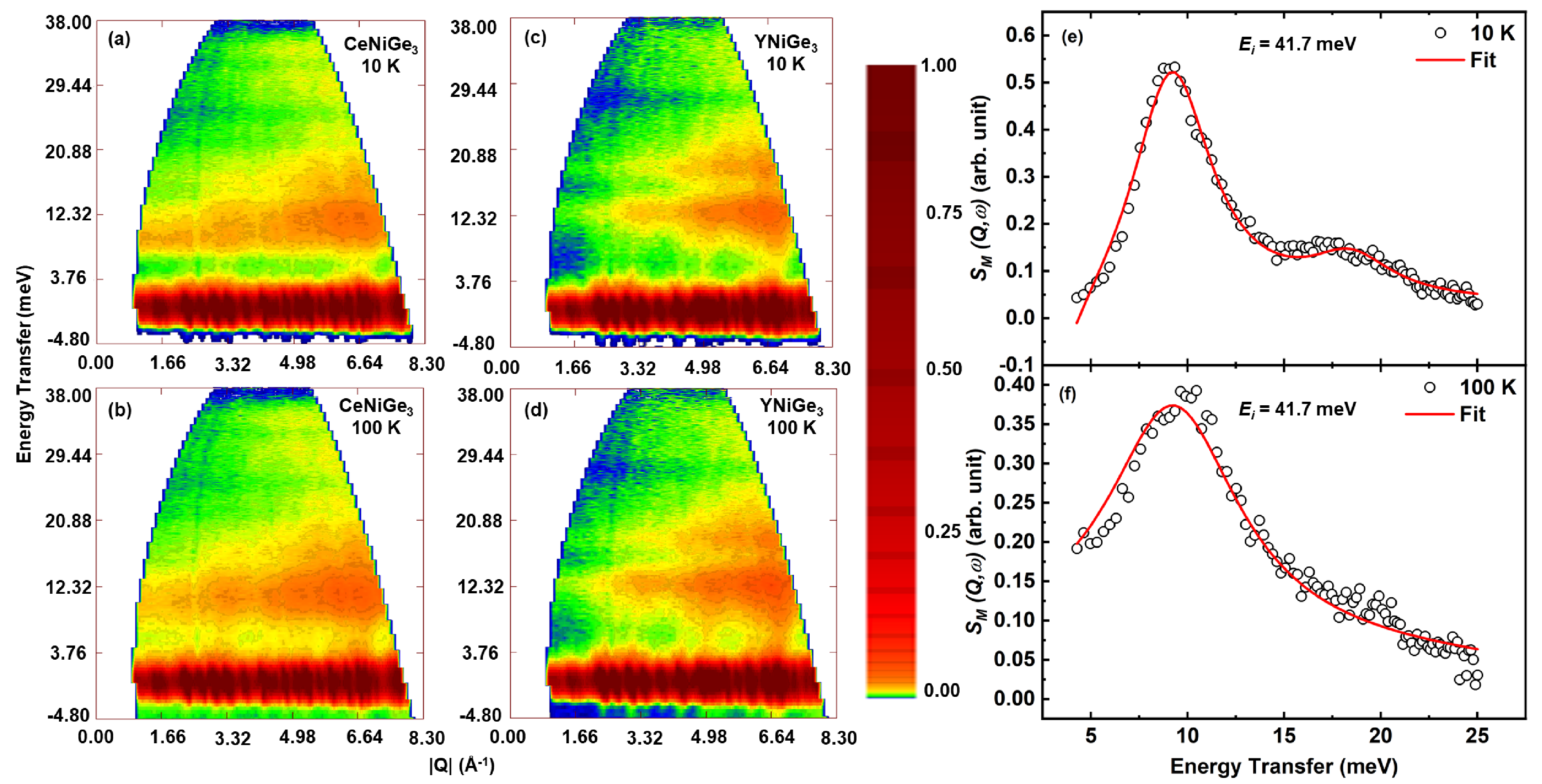}
    \caption{\label{ins data} Inelastic neutron scattering response, a color-coded contour map of the intensity, energy transfer $E$ vs momentum transfer $Q$ for CeNiGe$_3$ measured at (a) 10 and (c) 100 K, and for YNiGe$_3$ measured at (b) 10 and (d) 100 K, respectively. Magnetic scattering S$_M$(Q, $\omega$) vs energy transfer $E$ for CeNiGe$_3$ from 4 meV to 25 meV at (e) 10 K and (f) 100 K. The thick solid red lines represent the fit based on the CEF model using Eq. \ref{H_CEF}.}
\end{figure*}

\subsection {Neutron Diffraction}

In order to ascertain the characteristics of the long-range magnetic ordering of CeNiGe$_3$, we performed neutron powder diffraction experiments on a polycrystalline sample. The NPD data measured at 5.5 K can be indexed and refined assuming solely the structural nuclear contribution, however, the presence of additional magnetic Bragg peaks is found at 1.5 K. 

The Rietveld refinement profile of the 5.5~K data is shown in \figref{Fig3:neutron diff}(a), confirming that CeNiGe$_3$ crystallizes in the SmNiGe$_3$-type orthorhombic structure (space group: $Cmmm$) with lattice parameters $a$ = 21.7380(11) \AA, $b$ = 4.1170(2) \AA, and $c$ = 4.1511(2) \AA,  which are in good agreement with the previously reported values \cite{CNG3_poly,CNG3_SC,CTX3_SC}. The crystallographic parameters obtained from the refinement of NPD data of CeNiGe$_3$ are listed in Table \ref{tab:NPD}.

\begin{table}[b]
\centering
\caption{\label{tab:NPD} Crystallographic parameters of CeNiGe$_3$ obtained from the Rietveld refinement of the neutron powder diffraction data at 5.5~K (space group: $Cmmm$, No. 65). The estimated lattice parameters are $a$ = 21.7380(11)~\AA,   $b$ = 4.1170(2)~\AA ~and $c$ = 4.1511(2)~\AA.}
\begin{ruledtabular}
\begin{tabular}{lccccc}
\textrm{Atom}& \textrm{Wyckoff}& \textrm{x}& $y$ &  $z$& $B_{\rm iso}$(\AA$^2$)\\
\hspace{1.5cm} & position & \hspace{2cm} & \hspace{1cm} & \hspace{1cm} & \\
\hline
Ce & 4$j$ & 0.1679(3) & 0.0 & 0.5 &  0.013(3) \\
Ni & 4$i$ & 0.3916(1) & 0.0 & 0.0 &  0.024(4) \\
Ge1 & 4$i$ & 0.2839(2) & 0.0 & 0.0 &  0.021(5) \\
Ge2 & 4$i$ & 0.0559(2) & 0.0 & 0.0 &  0.022(5) \\
Ge3 & 4$j$ & 0.4435(2) & 0.0 & 0.5 &  0.019(6) \\
\end{tabular}
\end{ruledtabular}
\end{table}

A difference plot of the neutron diffraction patterns measured at 1.5~K and 5.5~K, depicted in \figref{Fig3:neutron diff}(b), clearly shows a number of magnetic Bragg peaks at 1.5 K. Fig. \ref{Fig:Mag_int} shows the temperature dependence of the integration of the intensity of the strongest magnetic peak which determines the magnetic transition temperature to be $T_{\rm N} \approx 5.5$~K. In order to understand the critical behavior we made an attempt to fit the integrated intensity by $I = I_0(1 - T /T_{\rm N} )^{2\beta}$, and found the value of critical exponent $\beta = 0.33(3)$ and $T_{\rm N} = 5.45(5)$~K. The fit of integrated intensity for $3.0~{\rm K} \leq T \leq T_{\rm N}$ is shown by a solid red curve in Fig. \ref{Fig:Mag_int}. The value of $\beta$ is much smaller than the mean-field value of 0.5, but it is close to that of a 3-dimensional Heisenberg antiferromagnet \cite{Blundell2001}.

All magnetic Bragg peaks in the magnetic only (difference) plot can be indexed by an incommensurate propagation vector {\bf k} = (0, 0.41, 1/2) using the {\bf k}-search program of FullProf suite \cite{fullprof}, which agrees very well with the previous CeNiGe$_3$ single crystal studies \cite{CNG3_singlecrystal}. Symmetry analysis was performed using the program BASIREPS \cite{basireps1,basireps2} for the Wyckoff position 4$j$ of the Ce ion in $\it {Cmmm}$ and the propagation vector {\bf k} = (0, 0.41, 1/2) and resulted in four two-dimensional irreducible representations (IRs); $\Gamma_1^2$, $\Gamma_2^2$, $\Gamma_3^2$ and $\Gamma_4^2$. Only IR  $\Gamma_3^2$ is able to refine the difference data containing the purely magnetic diffraction intensity as shown in \figref{Fig3:neutron diff}(b). The magnetic refinement profile ($R_{\rm mag} = 4.42 \%$) with this IR is shown in \figref{Fig3:neutron diff}(b). 
The incommensurate magnetic structure of CeNiGe$_3$ corresponds to a helicoid, as predicted earlier \cite{CNG3_singlecrystal}, but it has an elliptical envelope. Fig.~\ref{Fig:Mag_Struct} depicts this incommensurate magnetic structure. Within this helicoid, the magnetic moment of the Ce spins varies from 0.7 $\mu_\mathrm{B}$ to 1.07 $\mu_\mathrm{B}$ in the ellipsoid where the short axis is along the $a$-direction and the long axis along $b$. There is no component of the moment along c-axis, which is in agreement with the hard axis of the magnetization observed in the single crystal susceptibility and crystal field model discussed in section c. Ignoring the elliptical envelope by assuming a perfect helix leads to a drastically worsened refinement with $R_{\rm mag} = 11\%$.



\subsection {Inelastic Neutron Scattering}

The results of inelastic neutron scattering on CeNiGe$_3$ and its non-magnetic analog YNiGe$_3$ measured at 10 and 100 K are shown as color-coded intensity maps in \figref{ins data}(a),(c) and (b),(d), respectively. A careful comparison of the spectra reveals broad magnetic excitations in CeNiGe$_3$ for low momentum transfer values. At 10~K, there are two magnetic excitations, while at 100 K the linewidth of the high energy excitation is increased, see \figref{ins data}(e) and (f). Importantly, the energy value corresponding to excitations remained constant when the temperature varied from 10 to 100 K, suggesting that it originates from crystal electric field effects \cite{Cecuge3_muon}.\par

The magnetic contribution $S_M(Q,\omega)$ of CeNiGe$_3$ at low Q has been estimated by scaling using the YNiGe$_3$ INS data. We used YNiGe$_3$ high Q and low Q data to obtain the energy dependent scaling factor. The magnetic scattering of CeNiGe$_3$ at low Q was estimated by scaling its high Q data and using the energy dependent scale factor as  $S_M(Q,\omega)_{Low~Q} = $(Ce(Low Q) - Ce(High Q))$/$(Y(High Q)/Y(Low Q)). For a Ce$^{3+}$ ion in an orthorhombic point symmetry ($C_{2\nu}$), the Hamiltonian based on the point charge model, with $c$-axis ($c$$\parallel$$z$) as a quantization axis, can be written as 
\begin{equation}\label{H_CEF}
	H_{CEF} = B_2^0O_2^0 + B_2^2O_2^2 + B_4^0O_4^0 + B_4^2O_4^2 + B_4^4O_4^4
\end{equation}
where $B_m^n$ are CEF parameters while $O_m^n$ represents Stevens' operators \cite{stevens1952matrix}. The $B_m^n$ parameters can be estimated by fitting the experimental data, including INS and/or single-crystal susceptibility data. In this case, we used the parameters from Ref.~\cite{CNG3_SC} as an initial set of CEF parameters to fit the magnetic part of INS data. Also, we have used a Lorentzian line shape for the CEF excitations. To get a reliable set of CEF fitting parameters, we have simultaneously fitted the INS spectra at 10 K and 100 K, as depicted in Fig. \ref{ins data}. The obtained fitting parameters are tabulated in Table \ref{tab:cef}. As a result, we obtained slightly different CEF parameters compared to those reported in Ref.~\cite{CNG3_SC} and slightly different energy values for the first (9.17 meV) and second (18.42 meV) excited Kramers doublet. The CEF wave functions of the ground state, first excited state and second excited state of Ce$^{3+}$ in CeNiGe$_3$ are  
\begin{equation}
 \begin{split}
 \Psi_0 & = 0.8006\, | \pm \frac{1}{2}\rangle +0.5936\,|\mp \frac{3}{2} \rangle +0.0813\,|\pm \frac{5}{2} \rangle   \\
 \Psi_1 & = -0.5757 \,|\pm \frac{1}{2}\rangle +0.7997 \,|\mp \frac{3}{2} \rangle -0.1705 \, |\pm \frac{5}{2} \rangle \\ 
 \Psi_2 & = -0.1662 \,|\pm \frac{1}{2}\rangle +0.0897 \,|\mp \frac{3}{2} \rangle +0.9820 \, |\pm \frac{5}{2} \rangle. \nonumber
 \end{split}
\end{equation}

\begin{figure}
\centering
    \includegraphics[width=0.9\linewidth]{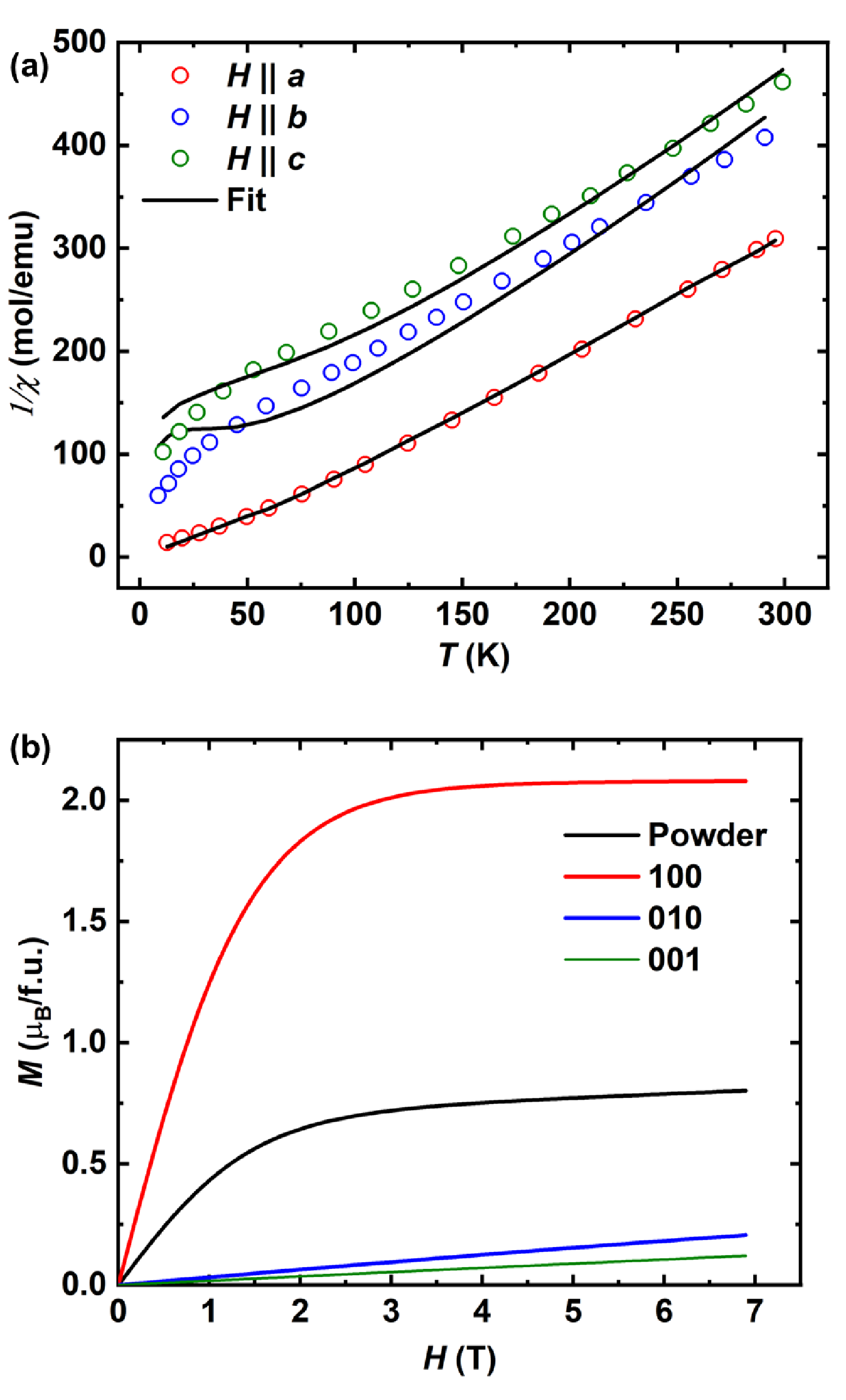}
    \caption{\label{sus_mag} (a) A comparison of CEF susceptibility (solid lines) with the susceptibility of single crystal CeNiGe$_3$ taken from Ref. \cite{CNG3_SC}. The $\chi_{\rm CEF}$ corresponds to the CEF parameters obtained from the fitting of the INS data by the CEF model [Eq. \ref{H_CEF}]. It is to be noted that in Ref.~\cite{CNG3_SC} the definition of the lattice parameters a and b is different (the long axis is b axis) than used in the present work (here the long axis is a axis). For the crystal field calculation the convection of Ref.~\cite{CNG3_SC} was used with $a//x$ (1 0 0), $b//y$ (0 1 0) and $c//z$ (0 0 1) where $x,y,z$ are the coordinates of the crystal field model.(b) Simulation of $M$ vs. $H$ at $T$ = 2 K for different directions and their average in the first quadrant. }
\end{figure}

Using the above information we could successfully reproduce the single crystal susceptibility data down to 50 K, as shown in \figref{sus_mag}(a). The CEF simulated magnetization curves [\figref{sus_mag}(b)] show that the value of $M$ is highest when the applied magnetic field is along the $a$ axis, suggesting that the $a$ axis is the easy axis for CeNiGe$_3$, which is also inferred from the magnetization measurements on single crystal sample \cite{CNG3_SC}. 
The polycrystalline average shows a value of 0.8~$\mu_\mathrm{B}$ at 2 K and 7~T, which is comparable to the experimental value of 0.89 $\mu_\mathrm{B}$ at 2~K and 7~T [see \figref{Fig1:charc}(c)]. It is worth mentioning that similar behavior has been observed earlier \cite{CNG3_SC, Durivault2003}. It is important to note that the definition of $a$ axis in Ref.~\cite{CNG3_SC} and in our CEF analysis, is the same as $b$ axis in our neutron diffraction analysis. Hence the easy axis of the magnetization and the observed direction of the moment (long moment along $b$ axis) from the neutron diffraction measurement are in agreement. 

\begin{table}
\centering
\caption{\label{tab:cef} Crystal electric field parameters ($B_n^m$) obtained from the analysis of inelastic neutron scattering data of CeNiGe$_3$. All parameters are in meV. For comparison we have also given CEF parameters from Ref.~\cite{CNG3_SC}}
\begin{ruledtabular}
\begin{tabular}{c c c}
CEF parameters & Present Work & From Ref.~\cite{CNG3_SC}\\
\hline
\textrm{$B_2^0$} & 0.8328 & 0.4619 \\
\textrm{$B_2^2$}& -0.9192 &  -0.9608 \\
 \textrm{$B_4^0$}& 0.0084 & 0.0103 \\
 \textrm{$B_4^2$}& 0.02809 & 0.04998 \\ 
 \textrm{$B_4^4$} & 0.01390 & 0.08789 \\
\end{tabular}
\end{ruledtabular}

\vspace{10 pt}

\end{table}

\subsection {Zero-field $\mu$SR}

\begin{figure}
\centering
    \includegraphics[width=\linewidth]{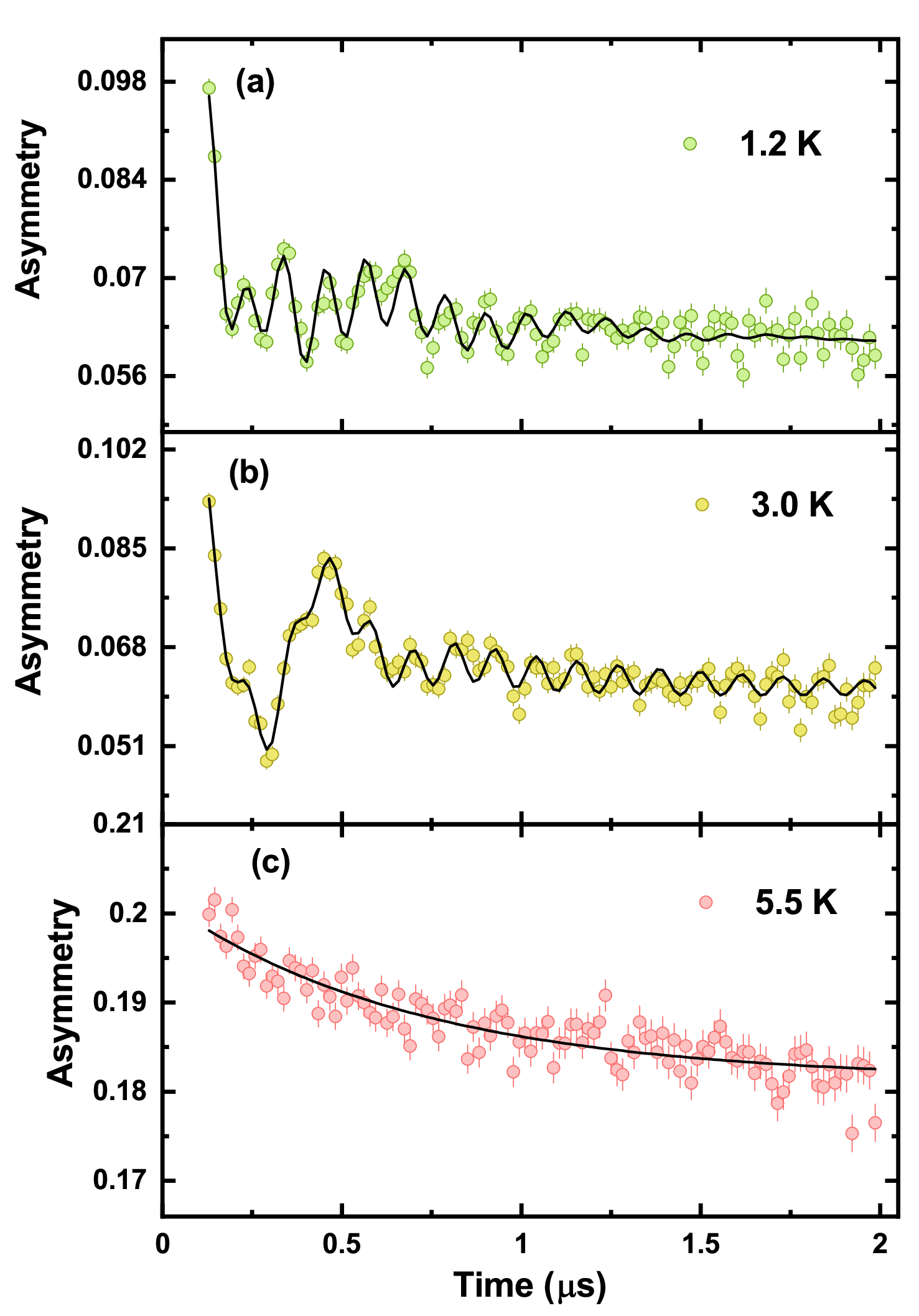}
    \caption{\label{Fig4:asymmetrymuon} ZF-$\mu$SR asymmetry spectra in the magnetic state at (a) $T= 1.2\,\text{K}$, (b) $T= 3.0\,\text{K}$ and (c) in the paramagnetic state, $T=5.5\,\text{K}$, with the solid black lines representing the respective fits using \equref{eq1:zf}. }
    \end{figure}
ZF-$\mu$SR measurements have been performed on polycrystalline CeNiGe$_3$ at regular temperature intervals from the lowest temperature (1.2~K) to above the AFM transition, $T_{\rm N}$. Figures~\ref{Fig4:asymmetrymuon}(a) and (b) depict the time-domain asymmetry spectra at 1.2~K and 3.0~K, respectively, where the observed oscillations in the ZF spectra suggest the presence of long-range magnetic ordering and a quasistatic internal magnetic field. The oscillatory feature in the spectra completely vanishes at 5.5 K (or above $T_{\rm N}$), and only exponentially decaying spectra are observed (\figref{Fig4:asymmetrymuon}(c)). The representative temperature evolution of the ZF-$\mu$SR asymmetry spectra is displayed in \figref{Fig4:asymmetrymuon}(a)-(c). To understand the nature of the local magnetic field and ordering in the ground state of CeNiGe$_3$, the ZF-$\mu$SR asymmetry spectra were fitted using the expression,
\begin{eqnarray}
    A(t) & = & \sum_{i=1}^N  A_i \cos(\gamma_{\mu}B_i t+ \phi) \exp\left(-\frac{1}{2}\sigma_{i}^2t^2\right) \nonumber \\
    &  & \hspace{3cm} + A_0 \exp(-\lambda t) + A_{bg}
     \label{eq1:zf}
\end{eqnarray}
where the first term, which features oscillatory Gaussian relaxation, includes the contribution of different magnetic-field components $B_i$ corresponding to different muon stopping sites, with initial asymmetries $A_i$, relaxation rates $\sigma_i$ and phase offset $\phi$ with $\gamma_{\mu}$ being the muon gyromagnetic ratio. The second term in the above expression is  an exponential decay with  relaxation rate $\lambda$ and asymmetry $A_0$, and the last term ($A_{bg}$)  accounts for the flat background contribution from the sample holder and cryostat. 

\begin{figure}
\centering
    \includegraphics[width=\linewidth]{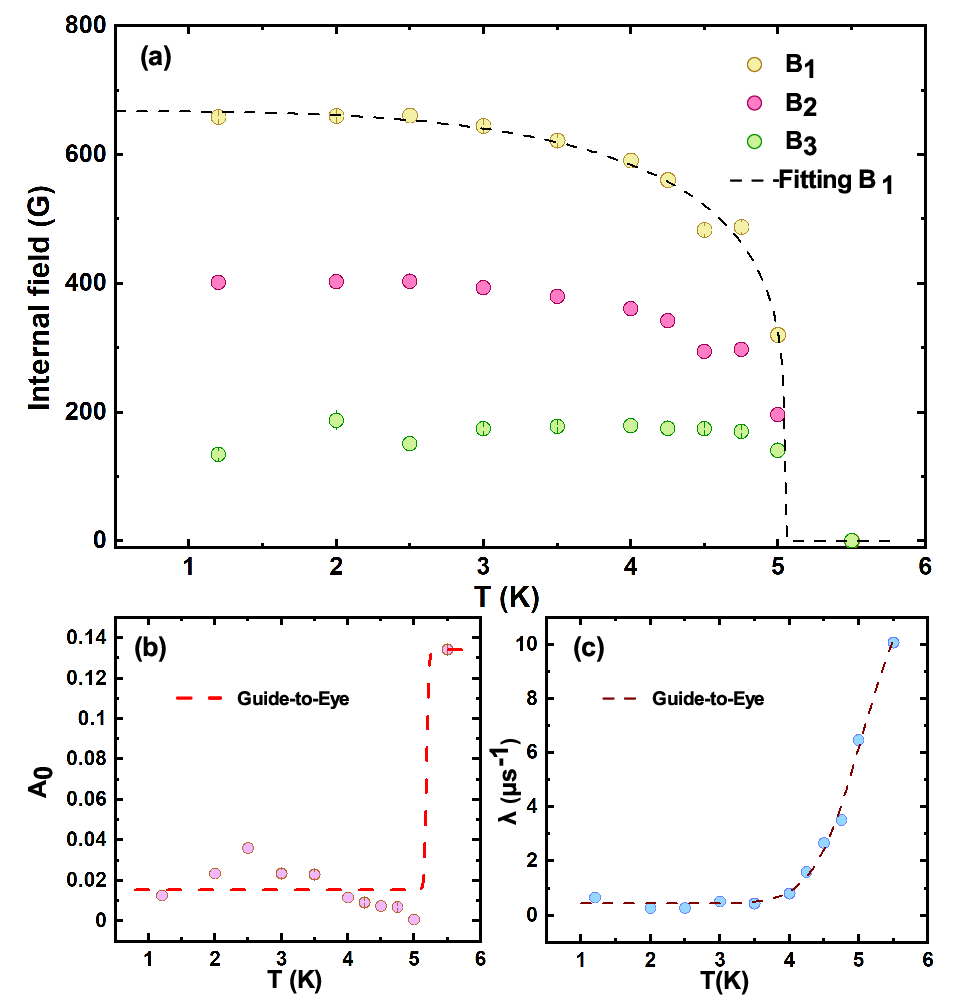}
    \caption{\label{Fig5:muonfreq} The temperature dependence of three distinct internal magnetic fields, $B_1$, $B_2$ and $B_3$. The dashed line corresponds to the fitting using \equref{eq2:bo}. (b) and (c) depict the asymmetry $A_0$  and $\lambda$ variation around the transition temperature, respectively, where dashed lines are guides-to-the-eye. }
    \end{figure}

Using $N = 3$ in the above equation describes the asymmetry spectra well below 5.5~K and over a short time scale (2.0 $\mu$s). However, it is to be noted from \figref{Fig4:asymmetrymuon}(a) that a large fraction of the initial asymmetry is missing at 1.2 K, which indicates the presence of a fast-relaxing signal involving larger internal fields.  These fields correspond to frequencies beyond the resolution of the MuSR spectrometer at ISIS  due to the pulse width (80ns) of the muon beam. For the fitting at 5.5~K, the first term in \equref{eq1:zf} does not contribute, and only the latter terms are considered. The temperature dependence of fitting parameters $B_i$, $A_0$ and $\lambda$ is illustrated in \figref{Fig5:muonfreq}. The background asymmetry parameter $A_{bg}$ = 0.065 remains temperature independent and is therefore kept constant throughout the temperature range for asymmetry fitting. The ratio between the second and first frequency components has been fixed to 0.61, determined from the value obtained in the low-temperature fit, which reduces the codependence of fitting parameters and yields accurate results. However, as can be observed in \figref{Fig5:muonfreq}(a) (and also fit to the asymmetry-time spectra near $T_{\rm N}$, not shown here) around the transition temperature, our model fails to completely describe the experimental data.  This indicates the presence of a more complicated local magnetic field distribution (i.e.\ one corresponding to more than three frequencies near $T_{\rm N}$). This distribution is only revealed at elevated temperatures, where high frequencies that were beyond the resolution limit at low $T$ now fall within the measurable frequency window. The details about the probable field distribution are discussed in next section.

Below $T_{\rm N}$, the asymmetry signal $A_0$ drops significantly from its high-temperature value, suggesting only a single magnetic phase transition, as shown in \figref{Fig5:muonfreq}(b). This is consistent with the observations from the neutron diffraction experiment and also depicts the dominance of the exponential function above 5.0~K, with a increase in $\lambda$ above $T_{\rm N}$, (see \figref{Fig5:muonfreq}(c)).  
 The temperature dependence of the first component of internal field $B_i$ is analysed using the relation \cite{bt_var}, 
\begin{equation}
    B (T) = B_0 \left[1-\left(\frac{T}{T_{\rm N,\mu SR}}\right)^{\alpha}\right]^{\beta}
    \label{eq2:bo}
\end{equation}
where $T_{\rm N, \mu SR}$ is the transition temperature from $\mu$SR and $\alpha$ and $\beta$ are the fitting parameters. The $B_1$ fitting of the data yields $\alpha$ = 3.7(9), $\beta$ = 0.22(8), $T_{\rm N, \mu SR}$ = 5.05(5) K and $B_0$ = 667(16) G. The $T_{\rm N, \mu SR}$ compliments the value obtained from our susceptibility (5.5 K) and heat capacity measurements (5.2 K). $\alpha$ and $\beta$ are the two numerical fitting parameters which depend on the type of magnetic interactions present in the system and its dimensionality \cite{Blundell2001, alpha}. However, in our case, the obtained beta value is only 0.22(8), significantly lower than expected for a mean-field-type interaction or 3-dimensional Heisenberg antiferromagnet \cite{Blundell2001}. The  probable reason for the discrepancy is insufficient data points close to  T\textsubscript{N}, where the rapid drop in internal fields makes it challenging to accurately determine the $\beta$ value in relation to the system's dimensionality or spin characteristics.


\subsection{Muon Stopping Sites}

We have carried out density functional theory (DFT) calculations using the plane-wave basis-set electronic structure code \textsc{castep} \cite{CASTEP} in order to determine the muon stopping sites \cite{muonsite}. Calculations were carried out within the generalized-gradient approximation (GGA) using the PBE functional \cite{PBE}. CeNiGe$_3$ crystallizes in the orthorhombic $Cmmm$ (No. 65) space group, see table-\ref{tab:NPD}. Structural relaxations were carried out on a supercell comprising $1\times2\times2$ conventional unit cells of CeNiGe$_3$ to suppress the unphysical interaction of the muon with its periodic images in plane-wave DFT.  We used a plane-wave cutoff energy of $700$ eV and $2\times 4 \times 4$ Monkhorst-Pack grid \cite{mpgrid1976} for Brillouin zone sampling, resulting in total energies that converge to within 1 meV per atom. The system was treated as non-spin-polarized in these calculations. These parameters resulted in DFT-optimized lattice constants that were within 1\% of experiment; these were therefore fixed at their experimental values for subsequent calculations.
 
\begin{figure}[htb]
	\centering
	\includegraphics[width=\columnwidth]{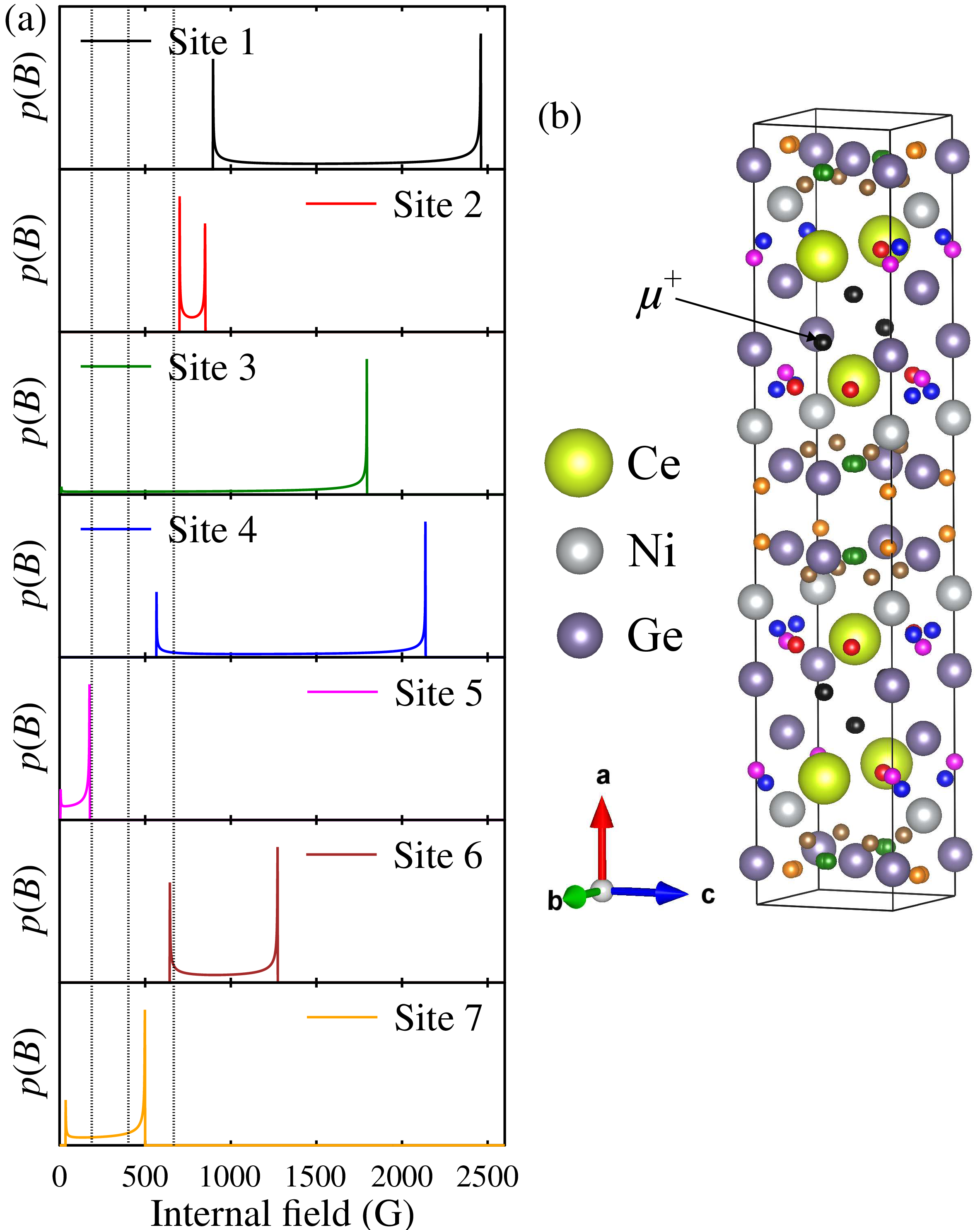}
	\caption{Field distribution at each of the muon sites in CeNiGe$_3$ due to the incommensurate magnetic structure. The three measured muon precession frequencies are indicated by the dotted vertical lines. (b) All of the distinct muon stopping sites obtained from structural relaxations using density functional theory. Sites are colored according to the scheme used in (a).}
	\label{fig:sites}
\end{figure}

Muon site calculations were carried out using the MuFinder program \cite{mufinder}. Initial structures comprising a muon (modelled as a light proton) and the CeNiGe$_3$ unit cell were generated by requiring the muon to be at least 0.5~\AA~away from each of the initial muon positions in the previously generated structures (including their symmetry equivalent positions) and at least 1.0~\AA~away from any of the atoms in the cell.  This resulted in 110 structures which were subsequently allowed to relax until the calculated forces on the atoms were all $<5\times 10^{-2}$ eV \AA$^{-1}$ and the total energy
and atomic positions converged to within $2\times10^{-5}$ eV per atom and $1\times10^{-3}$~\AA, respectively. These structural relaxations result in seven distinct muon stopping sites, whose properties are summarized in Table~\ref{tab:sites}. These sites are shown within the unit cell of CeNiGe$_3$ in Fig.~\ref{fig:sites}(b). Some initial positions resulted in muon sites with energies that were more than 1 eV higher than that of the lowest energy site; such sites are very unlikely to be realized in practice and were therefore discarded. 

In order to find the muon stopping sites the Ce $4f$ electrons were treated as valence electrons using the PBE functional. This approach yields values for the electron localization function (ELF) of up to 0.874 in the vicinity of Ce, with ELF = 1 representing perfect localization and ELF = 1/2 representing electron-gas-like pair probability. Applying a Hubbard U correction would likely increase the localization of the Ce $4f$ electrons. However in previous studies \cite{Franke2018} it was found that the Hubbard U does not have a significant effect on the resulting muon stopping sites. 

\begin{table}[htb]
	\caption{\label{tab:sites}%
		Muon stopping sites in CeNiGe$_3$ obtained from DFT using the structural relaxation method. Also shown are the energies of these sites, $E$, relative to the lowest energy site and the average dipolar magnetic field $B_\mathrm{avg}$ seen by a muon at this site.
	}
	\begin{ruledtabular}
		\begin{tabular}{lccr}
			\textrm{Site no.}&
			\textrm{Fractional coordinates}&
			$E$ \textrm{(eV)}&
			\textrm{$B_\mathrm{avg}$ (G)} \\
			\colrule
			1 & (0.276 0.002 0.494) & 0.000 & 1771 \\
			2 & (0.059 0.024 0.485) & 0.382 &  777 \\
			3 & (0.153 0.151 0.001) & 0.389 & 1143 \\
			4 & (0.165 0.500 0.296) & 0.427 & 1468 \\
			5 & (0.172 0.000 0.000) & 0.477 &  112 \\
			6 & (0.080 0.230 0.270) & 0.636 & 985 \\
			7 & (0.036 0.472 0.036) & 0.880 & 320 \\
		\end{tabular}
	\end{ruledtabular}
\end{table}

\subsection{Dipolar Fields}

Dipolar field calculations at the muon site are carried out using the MuESR Python library \cite{Bonfa2018}, which makes use of the method described in Ref.~\cite{Martin2016} to efficiently calculate the distribution of magnetic fields due to a helical magnetic structure. Here, we decompose the incommensurate magnetic structure of CeNiGe$_3$ into helices, and then use the fact that the dipolar field is linear in the magnetic moment to compute the dipolar fields arising from each of the constituent helices within this expansion. Corrections to the dipolar fields resulting from the muon-induced displacement of nearby magnetic ions are accounted for by using the approach described in Ref.~\cite{mufinder}, extended for use with incommensurate magnetic structures. The local magnetic field at the site is the sum of the dipolar, hyperfine, Lorentz and demagnetizing fields. In an antiferromagnetically ordered (or helical) state, the Lorentz and demagnetizing fields are zero. The hyperfine contribution to the magnetic field at the muon site can be difficult to determine, so we therefore assume that the dipolar field at the muon-stopping site is the dominant contribution to the local magnetic field experienced by the muon.

For an ellipitical magnetic structure $\boldsymbol{m}(\boldsymbol{r})$, whose moments vary in magnitude between $m_\mathrm{min}$ and $m_\mathrm{max}$, we can write this as the sum of two helices of opposite handedness
\begin{multline}
	\boldsymbol{m}(\boldsymbol{r}) = \Re \left[ \frac{1}{2}(m_\mathrm{max}+m_\mathrm{min}) (\mathbf{a}+i\sigma_\nu \mathbf{b})\exp(-2\pi i \boldsymbol{k} \cdot \boldsymbol{r}) \right.
	\\ \left. +  \frac{1}{2}(m_\mathrm{max}-m_\mathrm{min}) (\mathbf{a}-i\sigma_\nu\mathbf{b})\exp(-2\pi i \boldsymbol{k} \cdot \boldsymbol{r}) \right],
\end{multline}
where $\mathbf{a}=(0,1,0)$ and $\mathbf{b}=(1,0,0)$ are unit vectors describing the plane of rotation of the helix and the propagation vector $\boldsymbol{k}=(0,0.41,0.5)$. The parameter $\sigma_\nu=\pm1$, describes the handedness of the elliptical helix; $\sigma_\nu=1$ for a right-handed helix, whereas $\sigma_\nu=-1$ gives a left-handed helix. For the incommensurate magnetic ground state of CeNiGe$_3$, we have $m_\mathrm{min}=0.7\mu_\mathrm{B}$ and $m_\mathrm{max}=1.07\mu_\mathrm{B}$. Obtaining the correct handedness for each of the helices requires $\sigma_\mathrm{Ce1}=\sigma_\mathrm{Ce3}=-1$ and $\sigma_\mathrm{Ce2}=\sigma_\mathrm{Ce4}=1$.

The dipolar magnetic field at each muon site was calculated using this magnetic structure. Note that, due to the fact that the magnetic sturcture is incommensurate, each crystallographically distinct muon site sees a distribution of dipolar fields. These are continuous field distributions characterized by lower and upper cut-off fields, as also seen in helimagnets such as MnSi~\cite{Reotier2016} and MnGe~\cite{Martin2016}. We also calculate the average dipolar magnetic field experienced by a muon at each of these site, and list these in the final column in Tab.~\ref{tab:sites}. 

It is  clear that none of the calculated field distributions convincingly reproduce the internal magnetic fields observed in experiment. Some of them do exhibit peaks close to  the measured fields, for example, the lower peaks for sites 2, 4 and 6, or the the upper peaks for sites 5 and 7. However, it is notable that the lowest-energy site, which is expected to be the most likely to be occupied, results in the muon experiencing much larger dipolar magnetic fields than the internal fields measured in experiment. 
One possible reason for a discrepancy between the calculated and measured fields is the omission of the contact hyperfine field at the muon site in these calculations. This could, in principle, act to reduce the total magnetic field at the muon site, thereby bringing these into slightly closer agreement with those seen in experiment. However, this effect would still leave the question of how to reconcile the large number of rather different muons sites with the three dominant oscillation frequencies observed in the data.

However, we note that for fields above $\approx 500$~G  the resolution limitation that results from the width of the ISIS muon pulse leads to a reduction in asymmetry. This causes muon sites in magnetic fields $> 500$~G to contribute to the missing asymmetry. Given that we observe a sizable fraction of missing asymmetry it is highly likely that this mechanism occurs here. We therefore have a picture in which overlapping low-field (and relatively high-energy) muon sites such as 5 and 7 likely contribute to the resultant spectral weight to give the two low-field oscillations centred at fields $B_{2}$ and $B_{3}$. In contrast, to obtain the oscillation centred at the large field $B_{1}$, sites such as 2, 4 and/or 6  likely contribute, but also contribute to the missing portion of asymmetry. This  broadly accounts for the muon spectra within the assumed model of the magnetic structure, although we note that this requires the realization of a several (i.e.\ two or three) sets of muon sites, some  sitting at a significant fraction of an eV above the lowest-energy muon site, whose field profiles overlap to give the three dominant fields found in our analysis. The realization of high-energy muon sites depends on the capture cross section of these states during the muon stopping process, about which we have limited information, so it will be interesting to see how this case compares to the sites realized in similar materials in future studies.


\section{Summary and Conclusions}

We have investigated the magnetic properties of a centrosymmetric orthorhombic SmNiGe$_3$-type structured compound CeNiGe$_3$ in detail through macroscopic and microscopic measurements, including $\chi(T)$, $M(H)$, $C_{\rm p}(T)$, NPD, INS, and $\mu$SR. The $\chi(T)$, $M(H)$ and $C_{\rm p}(T)$ measurements confirmed the reported antiferromagnetic magnetic ordering ($T_{\rm N} \approx 5.2$~K) with meta-magnetic transitions in our sample. Our $C_{\rm p}(T)$ data do not support a heavy fermion behavior in this compound. Our neutron diffraction study reveals a helicoidal spin structure with a single incommensurate propagation vector \textbf{k} = (0, 0.41, 1/2) at 1.5~K\@. This agrees with neutron diffraction studies on a single crystalline sample; however, it contradicts the previous report on a polycrystalline sample, which reports the coexistence of two different magnetic structures \cite{CNG3_poly}.  The presence of a single incommensurate magnetic structure is consistent with the NQR study, which is highly sensitive to spatial inhomogeneities around nuclei to discern the transitions between the two possible magnetic structures \cite{nqr1, nqr2}. The improved statistics and the use of difference data in the NPD refinement clearly ascertained an elliptical envelope of the helicoid describing the magnetic moment propagation. The INS data of CeNiGe$_3$ were fitted using a CEF model, revealing two CEF excitations with energies of 9.17 meV and 18.42 meV. The observation of only two excitations suggests the absence of CEF-phonon coupling, which has been observed in CeCuAl$_3$, CeAuAl$_3$ and CeCuGa$_3$ \cite{cef_coupling, Ceramak2019, Cecuge3_muon}. The obtained CEF parameters from the fitting were used to evaluate the ground state wave function and energy levels. Our fitting parameters differ slightly from the CEF parameters that were previously reported and were estimated solely from the single-crystalline susceptibility measurements. The agreement of our parameters with both the INS and the single-crystal susceptibility data makes the CEF parameters we obtained more reliable.  Furthermore, $\mu$SR measurements indicate the long-range magnetic ordering with three oscillation frequencies, whose temperature dependence also suggests a single transition at 5.05(5)~K ($T_{\rm N, \mu SR}$). 
\\
\section{Acknowledgments}

A.K. thanks the Indian Nanomission for a post-doctoral fellowship. R.K. acknowledges CSIR, India, for SRF fellowship (grant number: 09/733(0263/2019-EMR-I)). D.T.A. and V.K.A. acknowledge financial assistance from CMPC-STFC grant number CMPC-09108. D.T.A. thanks the Royal Society of London for the Newton Advanced fellowship funding between the UK and China and the International Exchange funding between the U.K. and Japan. D.T.A. also thanks EPSRC UK (Grant No. EP/W00562X/1) for funding and the CAS for PIFI fellowship.  Computing resources are provided by the STFC Scientific Computing Department's SCARF cluster. B.\,M.\,H. and TL acknowledge funding from EPSRC (UK) under Grant No. EP/N024028/1, and B.\,M.\,H. acknowledges funding from UK Research and Innovation (UKRI) under the UK government's UKRI funding guarantee (Grant No. EP/X025861/1). We thank ILL for beam time on IN4  Experiment No: 4-01-953, and on D1B, 5-31-2718 \cite{data_ref} and ISIS Facility for beam time RB1120500 \cite{data_ref1}. Data from this study will be made available via e-mail request to D.T.A.

\end{document}